# Monochromatic operation of the SASE3 soft X-ray beamline at European XFEL

N. Gerasimova[*], D. La Civita, L. Samoylova, M. Vannoni, R. Villanueva, D. Hickin, R. Carley, R. Gort, B. Van Kuiken, P. Miedema, L. Le Guyarder, L. Mercadier, G. Mercurio, J. Schlappa, M. Teichman, A. Yaroslavtsev, H. Sinn and A. Scherz

European XFEL, Holzkoppel 4, 22869 Schenefeld, Germany

* E-mail: natalia.gerasimova@xfel.eu

**Synopsis**   The SASE3 soft X-ray beamline at the European XFEL is equipped with a grating monochromator. The design and current performance of the monochromator are discussed.

**Abstract**   The SASE3 soft X-ray beamline at the European XFEL has been designed and built to provide experiments with pink or monochromatic beam in the photon energy range 250 eV - 3000 eV. Here, we focus on the monochromatic operation of the SASE3 beamline and report on design and performance of the SASE3 grating monochromator. The unique capability of an FEL source to produce short femtosecond pulses of high degree of coherence challenges the monochromator design by a demand to control both photon energy and temporal resolution. The aim to transport close to transform-limited pulses poses very high demands on the optics quality, in particular on the grating. The current realization of the SASE3 monochromator is discussed in comparison with optimal design performance. Presently, the monochromator operates with two gratings: the low-resolution grating is optimized for time-resolved experiments and allows for moderate resolving power of about 2000 - 5000 along with pulse stretching of few to few tens of femtoseconds RMS, and the high-resolution grating reaches resolving power of 10000 at a cost of larger pulse stretching.

## 1. Introduction

The SASE3 beamline at the European XFEL facility operates in the soft X-ray range from 250 eV to 3 keV [Altarelli *et al.*, 2006; Tschentscher *et al.*, 2017]. The undulators of the SASE3 beamline produce femtosecond pulses of intense highly coherent radiation that are transported to diverse experiments performed either at the SQS (Small Quantum Systems) instrument or at the SCS (Spectroscopy and Coherent Scattering) instrument; the third soft X-ray instrument SXP (Soft X-ray Port) is currently under construction. A considerable number of experiments conducted at the SASE3 beamline, in particular those applying spectroscopic techniques, demand for photon energy resolution substantially better than the typical 0.2 - 1% bandwidth of SASE (Self-Amplified Spontaneous

Emission) radiation. In order to reduce photon energy bandwidth at the experiments, the SASE3 beamline is equipped with a grating monochromator. The SASE3 beamline, along with the monochromator, provides beam for experiments since the end of 2018. Here, we present the design of the SASE3 monochromator, its status as of 2021, and the performance achieved in the first three years of operation.

The design of a beamline at an FEL source is driven by the aim not to compromise the unique properties of the FEL pulses by the beam transport. The high degree of coherence of the FEL pulses entails a goal to enable close to diffraction-limited beam transport. This goal sets tight demands on the quality of optical elements. Furthermore, the European XFEL produces trains of pulses of millijoule intensity level at a high intra-train repetition rate of up to 4.5 MHz, so that care should be taken to avoid damage to the optics. To reduce risk of damage, in particular critical in the soft X-ray range, where the interaction of radiation with matter results in very high densities of excitations, an increase in the footprint on optics and a choice of low Z materials for coating of reflecting optical elements in order to reduce absorbed dose are expedient. To allow large footprints of 4σ of a Gaussian-like beam profile, very long beamlines with very long optical elements are envisaged. Another property of FEL radiation is the ultrashort pulse duration of few to few tens of femtoseconds, which is important for a wide range of time-resolved experiments. The goal to minimize deterioration of temporal resolution by the monochromator sets additional demands on the grating of the monochromator. In contrast to a synchrotron beamline, where the monochromator design aims for the best compromise between photon energy resolution and transmission, in the case of an FEL source, pulse stretching by the grating should be taken into account as well. Since an amount of spectroscopic experiments demand for the best compromise between temporal and photon energy resolution, another aim is to minimize the time-bandwidth product of transported pulses. In order to minimize the time-bandwidth product and to transport close to transform-limited pulses, aberrations by optics should be minimized ideally down to the instrument response function of the ideal grating. This leads to very tight demands on the grating quality.

The design of the SASE3 monochromator is based on a combination of an elliptical mirror and a VLS (variable line spacing) grating. Reducing the number of involved optical elements to two allows to minimize the effect of figure error of optics on performance. In order to minimize the pulse stretching by the grating, a line density of 50 l/mm was chosen as the lowest possible to be produced by the start of operation of the European XFEL facility. The design foresees achieving > 4σ of Gaussian-like beam profile transmission in the substantial part of the photon energy range of SASE3 operation. However, achieving the design parameters for a 500 mm long VLS grating was outside of production capabilities. Therefore, a 120 mm long 50 l/mm grating, allowing for moderate photon energy resolution along with minimal pulse stretching, was produced and installed by the start of operation. In 2021, the beamline has been complemented by a 120 mm long 150 l/mm grating, allowing for higher photon energy resolution at a cost of larger pulse stretching. The performance of the two gratings in operation is discussed in the present work in comparison with the optimal design performance.

## 2. Layout of the SASE3 monochromator

The concept of the SASE3 monochromator was introduced in [Sinn *et al.*, 2012]. The layout of the SASE3 beamline is presented in Fig. 1; details on the optical elements are summarized in Table 1 and Table 2. The offset mirrors (M1 and M2) and the distribution mirrors (M5 and M6) deflect the beam in the horizontal plane and serve for the beam transport to the instruments: switching from the SQS instrument to the SCS instrument is done by inserting mirror M5, and from the SQS instrument to the SXP instrument - by inserting mirror M6. The adaptive mirror M2 allows for the horizontal focusing of the beam. The SASE3 monochromator itself operates in the vertical plane and consists of two optical elements: a meridianally focusing premirror M3 and a plane VLS grating. The premirror of elliptic cylinder shape focuses the source point, located at the exit of undulator system, onto the exit slit; the VLS ruling of the grating compensates for aberrations arising from the operation of the grating in the convergent beam. The monochromator operates at a fixed included angle; the two premirrors: LE (low energy) premirror, operating at a grazing angle of 20 mrad, and HE (high energy) premirror, operating at a grazing angle of 9 mrad, allow to cover the photon energy range 250 - 3000 eV. Change of the photon energy is done by rotation of the grating. The internal diffraction orders are used. Finally, there is a possibility to use a plane mirror M4 instead of a grating so that a pink beam of the same focusing properties as the monochromatic beam is transported to the experiments.

All the optical elements of the SASE3 beamline are made of single crystal silicon and coated with a 50 to 65 nm layer of B4C. The roughness of the substrates was specified as < 0.2 nm RMS, the tangential slope error of the mirrors as < 50 nrad RMS, the height error of the mirrors as < 3 / 2 nm for the vertically/horizontally deflecting optics. In order to reduce thermal deformations, the optical elements were designed to be cooled with InGa eutectic bath [Sinn *et al.*, 2012; La Civita *et al.*, 2014]. Up to now the cooling is implemented for the horizontally deflecting mirrors.

The key element of the monochromator is the grating. The design of the monochromator, based on two optical elements only, allows to minimize the number of involved optical elements and thus reduce wavefront deformation due to figure error introduced by every optical element. On the other hand, such a design demands for very high accuracy on the VLS parameters. In spite of existing advanced technologies of VLS grating production, there are two limiting particularities in the case of SASE3 grating. Firstly, to allow transmission > 4σ of Gaussian-like beam, the length of the grating was specified to be 500 mm, which is unusually long for VLS gratings. Importantly, the accuracies to be set on the VLS parameters increase drastically with the length of the grating. Secondly, in order to prevent damage on the optical elements and on the exit slits, the design of the beamline foresees very long distances: the distance from the nominal source to the grating comprises 301 m, the distance from the grating to the exit slit – 99 m. Such long focal distances result in very low values of VLS parameters. The low values of VLS parameters should not challenge grating production based on mechanical ruling; however, the set-ups for holographic grating production are oriented for larger values of VLS parameters. As a result, the design 500 mm long 50 l/mm grating happened to be beyond the production capabilities to be ready by the start of operation. As a working solution, a shorter, 120 mm long 50 l/mm grating was produced holographically by Horiba Jobin-Yvon. To allow for higher photon energy resolution, the beamline has been complemented recently by the 120 mm long 150 l/mm grating. The characteristics of these gratings are listed in Table 2. Below we discuss the expected performance of design grating and compare it with the performance of the two gratings in operation.

The monochromator operates in both fixed photon energy mode and scanning mode. For the scanning over photon energies, synchronization of the monochromator with the undulator system is applied. In addition, the spectrometer mode of operation of the SASE3 beamline is regularly utilized for spectral diagnostics of the FEL beam. The spectrometer mode of operation is realized by introducing the YAG:Ce crystal into the focal plane of the monochromator. The YAG:Ce crystal converts X-ray photons into optical luminescence, which is registered in turn by a CCD coupled with an MCP [Koch *et al.*, 2019]. The resulting 2D images represent the spectral distributions in the focal plane of the monochromator.

### 3. Methods

The performance of the SASE3 monochromator has been evaluated analytically over the SASE3 operational range. To estimate the beam footprint on the optics, and the consequent geometrical cut in vertical direction due to limited grating length, the FEL beam was approximated by a Gaussian-like beam, the beam divergence was estimated according to [Sinn *et al*., 2011], based on the expected SASE beam properties in the case of low electron beam charge of 20 pC presented in [Schneidmiller & Yurkov, 2011]. The resulting shape of the footprint on the grating (cut Gaussian-like beam, see inserts in Fig. 4(a)) was used to estimate the pulse stretching: $t=(M*k*\lambda*F)/c$, where $M$ is the diffraction order, $k$ is the line density, $\lambda$ is the photon wavelength, $c$ is the speed of light, $F$ is the beam footprint on grating. The corresponding beam profile after the grating was used to retrieve the monochromatic instrument response function (IRF) of ideal grating in the exit slit plane by Fourier transform. To obtain the IRF of a real grating, the IRF of an ideal grating should be convoluted with a function representing aberrations. Aberrations due to slope error on optics and due to miscut of the VLS parameters were taken into account. A simplified formula, assuming Gaussian shapes of the IRF and of the aberrations, was used to estimate the FWHM (full width at half maximum) of the IRF of real grating: $irf=\sqrt{(irf_{gr}^2+a_{slope}^2+a_{vls}^2)}$, *where* $irf_{gr}$ is the FWHM of the IRF of ideal grating given by beam footprint of the grating, $a_{slope}$ is the aberration induced by slope error, and $a_{vls}$ is the aberration induced by the VLS miscut. The aberration induced by slope error was estimated as a combination of aberration due to slope error on the premirror $a_{slope\_pm}=\sigma_{slope\_pm}*2/cff$ and aberration due to slope error on the grating $a_{slope\_gr}=\sigma_{slope\_gr}*(1+cff)/cff$, where $\sigma_{slope\_pm}$ and $\sigma_{slope\_gr}$ are the slope errors on the premirror and on the grating respectively, $cff=cos(\beta)/cos(\alpha)$ is the fixed focus constant, $\alpha$ is angle of incidence on the grating and $\beta$ is angle of diffraction. The resolution relates to the FWHM of IRF: $r=irf*dE$, where $dE$ is dispersion. The target VLS parameters and aberrations induced by miscut of the VLS parameters have been estimated analytically using the geometrical theory of diffraction grating [Noda *et al.*, 1974]. To estimate tolerances on the VLS parameters, aberrations induced by each VLS term have been set equal to half of the FWHM of IRF of the grating with ideal VLS. Such a condition would result in a resolution reduced by ~ 11% in the case of a Gaussian profile. The details can be found in [Gerasimova, 2018], where the study on the 50 l/mm grating is reported.

To confirm analytical estimations of IRF and resolution, wavefront propagation simulations have been applied for several working points. The WaveProperGator (WPG) framework [Samoylova *et al.*, 2016] was used; the source was modelled as a Gaussian-like beam. When departing from optimal conditions (e.g. due to imperfections of grating or due to defocus), the shape of the IRF becomes ambiguous, so the reduction of main peak intensity was used as a quantitative measure of resolution deterioration. The wavefront propagation simulations were used for the investigation of longitudinal focusing properties, such as dependence of resolution on premirror focusing distance in the case of a grating with residual curvature.

For estimating the time-bandwidth product, which is a product of FWHM pulse duration and FWHM photon energy bandwidth after the monochromator, the pulse duration after the monochromator was assumed to be equal to the pulse stretching by the grating, and the FEL pulse duration itself was neglected. This approximation is valid for the cases in which the pulse stretching is longer than FEL pulse duration at a fixed photon energy; the latter is typically of the order of 10 fs FWHM and below, as estimated by spectral correlations technique.

The transmission of the beamline has been estimated by combining (i) geometrical cut by the optics, (ii) grating efficiency and mirrors reflectivity, and (iii) spectral transmission through the exit slit. Grating efficiencies have been calculated using "Reflec" code [Schäfers & Krumrey, 1996], which is part of the "Ray" package [Schäfers, 1996] developed at the Berliner Elektronenspeicherring-Gesellschaft für Synchrotronstrahlung (BESSY). To estimate the spectral transmission through the exit slit, the exit slit was assumed to be centred to the maximum of the Gaussian-like spectrum of 1% bandwidth. In practice, the average FEL spectra differ from Gaussian shape and could have various bandwidth (from 0.2 - 0.5% of theoretical SASE bandwidth to > 1%). To compare measurements and estimations, the SASE spectrum was measured in every experimental point, and the spectral transmission through the exit slit of measured spectrum was normalized to the spectral transmission of the Gaussian-like spectrum of 1% bandwidth. The transmission was measured as the ratio of intensities, registered by two XGM (X-ray gas monitor) detectors [Maltezopoulos *et al.*, 2019] located before the first beamline mirror M1 and after the SCS exit slit.

The absorption resonances at Ne K-edge were used for the monochromator calibration and for the photon energy resolution measurements. The measurements were done in transmission: the 15 m long gas attenuator [Villanueva *et al.*, 2022] located upstream of the monochromator was filled with Ne gas; the pressure of Ne was in the range 0.005 - 0.03 mbar, depending on the monochromator configuration; the beamline was operating in the spectrometer mode so that the single shot spectra were acquired in the focal plane of the monochromator by a CCD gated with an MCP. The spectra were taken at 10 Hz, the ensemble of 5000 - 10000 single shot spectral distributions were averaged, and the resulting spectrum was normalized by an average FEL spectrum. Resulting transmission spectra were fit with a function resulting from a convolution of the modelled Ne K-edge transmission spectrum with the Gaussian line representing the IRF of the monochromator. To model the transmission in 1s excitation region of Ne, the absorption cross-section was modelled according to [De Fanis *et al.*, 2002] by Lorentzian lines of natural lifetime width $\Gamma$ = 240 meV, located at 867.12 eV ($1s^{-1}3p$), 868.69 eV ($1s^{-1}4p$), 869.27 eV ($1s^{-1}5p$), etc. The photon energy resolution was estimated as a FWHM of IRF of the monochromator resulting from the fit.

## 4. Performance of the SASE3 monochromator

### 4.1. Photon energy resolution and temporal resolution: design and limitations

The photon energy resolution was estimated analytically over the SASE3 operation photon energy range, as described in Section 3. To confirm analytical estimations, wavefront propagation simulations were performed at several working points.

The influence of figure error on photon energy resolution in the case of the 500 mm long 50 l/mm design grating operating in 1st diffraction order is presented in Fig. 2(a, b). The 50 nrad RMS slope error was specified for the premirrors and for the design grating; this value is presently at the limit of production and metrology capabilities. Analytical estimations show, that a combination of aberrations due to 50 nrad slope error on the premirror and aberrations due to 50 nrad slope error on the grating results in a slope-error-limited resolution dominating over the resolution of the ideal grating (Fig. 2a). The high energy range is more strongly influenced by the slope error, so that a deterioration of the overall resolution due to slope error is more substantial at higher photon energy ranges of operation of each premirror. The influence of the magnitude of figure error on the resolution of the 500 mm long 50 l/mm grating was also investigated by wavefront propagation simulations. In contrast to the analytical estimations, where the slope error introduces an angular spread of deflected rays with a consequent focus deterioration, in the wavefront propagation simulations the height error is introduced as a phase screen resulting in wavefront distortion of coherent beam and in consequent focus deterioration. The results of wavefront propagation show lower influence of figure error compared to analytical estimations: as it can be seen in Fig. 2(b), in the case of 1500 eV, the 3 nm height error corresponding to 40 nrad slope error on the grating would only lead to a small reduction of resolution compared to ideal slope on grating.

The estimations of resolution of a 120 mm long 50 l/mm grating operational in 1st diffraction order are presented in Fig. 2(c). Due to shorter grating length compared to a 500 mm long design grating, less grooves are illuminated, resulting in broader IRF and lower ideal resolution. Consequently, the worse slope error of 200 nrad RMS leads to the similar relative deterioration of resolution in the case of a 120 mm long grating as 50 nrad RMS slope error in the case of a 500 mm long grating. The influence of VLS miscut on the resolution was studied as well. The optimal VLS $b2$ for the 50 l/mm grating is $1.53*10^{-8}$ l/mm$^3$; the lowest VLS $b2$ achieved in the production of a 120 mm long 50 l/mm grating is $-1.2*10^{-6}$ l/mm$^3$. The aberrations induced by this miscut are inferior to the ideal resolution and do not limit the resolution of the 120 mm long 50 l/mm grating operational in 1st diffraction order (Fig. 2(c)). However, in 2nd diffraction order this VLS $b2$ miscut becomes already a limiting factor for the resolution of this grating.

Let us discuss the VLS parameters, which limit the production of design grating. The tolerances on the VLS parameters were estimated as described in Section 3 by setting the VLS induced aberration equal to half of FWHM of IRF of the grating with ideal VLS. The resulting tolerances on the VLS $b2$ parameter are presented in Fig. 2(d). Important to underline is a strong dependence of the illuminated grating length on the magnitude of aberrations induced by the VLS miscut: the aberrations due to a miscut of $b1$ would be proportional to the first power of the illuminated grating length, $b2$ – to the second power of the grating length, $b3$ – to the third power of the grating length, etc. In addition, the FWHM of IRF of ideal grating is inversely proportional to the grating length demanding for smaller aberrations for longer gratings. Therefore, the tolerance to be set on the VLS $b2$ parameter would be approximately proportional to the third power of the illuminated grating length. As shown in Fig. 2(d), if the full grating would be illuminated, the tolerances on the VLS $b2$ for a 500 mm long grating are almost by two orders of magnitude tighter than the tolerances for a 120 mm long grating. However, due to photon energy-dependent FEL beam divergence and due to the monochromator geometry, the 500 mm long grating is under-illuminated at higher photon energy ranges of operation of each premirror (Fig. 4(a)). When only partial illumination of the grating, corresponding to FWHM of the beam footprint, is taken into account to estimate aberrations due to the VLS parameters, the tolerances in the case of 500 mm long grating become less tight.

The strong dependence of aberrations induced by a miscut of VLS parameters on grating length discussed above is important to understand limitations set by grating production capabilities on reasonable grating length. If increasing the length of an ideal grating would improve the resolution with increase in illuminated part of the grating, the slope error would stop this improvement at some point when resolution would become slope-error-limited, and the VLS miscut would deteriorate the resolution increasingly with increase in the grating length as soon as the resolution becomes VLS-miscut-limited. This tendency can be seen in Fig. 3(a), where the analytically estimated resolution of 50 l/mm grating with 50 nrad slope error operating in 1$^{st}$ diffraction order is presented for the cases of ideal VLS and of VLS $b2 = -1.2*10^{-6}$ l/mm$^3$. In the latter case, the best resolution would be achieved at a grating length of about 130 mm and would deteriorate with further increase in the grating length, e.g. at 400 mm grating length, the resolution would be spoiled by a factor of 5 compared to optimal resolution at 130 mm length. For the 150 l/mm 120 mm long grating, produced a few years later after the 50 l/mm 120 mm long grating, the tighter VLS $b2 = -8.7*10^{-8}$ l/mm$^3$ has been achieved. Such a small VLS miscut does not introduce any impact on resolution for the operation in 1$^{st}$ diffraction order, and, assuming the longer grating with such VLS parameters could be produced, it would be possible to increase the grating length to 250 mm without resolution deterioration.

Another effect accompanying an increase in the grating length is an increase in the pulse stretching. The grating introduces pulse stretching due to the pulse front tilt resulting from difference between angle of incidence and angle of diffraction. Since this difference becomes more pronounced at lower photon energies, pulse stretching would be more pronounced at lower photon energies. Pulse stretching depends on the size of illuminated part of the grating: the longer illuminated part of the grating, the longer the pulse stretching. The illuminated part of the grating could be controlled by the aperture introduced before the grating. The dependence of pulse stretching by a 50 l/mm grating on the aperture size is presented in Fig. 3(b) for operation at 290 eV, 500 eV, and 800 eV. The stars correspond to the 120 mm length of the grating. Importantly, while the pulse stretching is given by geometry only and does not depend on the quality of the grating, the resolution degrades compared to best possible resolution with increase in the grating length. This results in an increase in the time-bandwidth product with the increase of the length of a real grating, starting from the point at which aberrations start to influence the resolution.

Let us discuss the time-bandwidth product in detail. As was pointed out, one of design goals was to minimize time-bandwidth product in order to provide experiments with the best compromise between temporal and photon energy resolution. First of all, in the case of ideal optics, the time-bandwidth product depends on the shape of the beam footprint on the grating only. It is well known that time-bandwidth product is minimal for a Gaussian-like beam, reaching ~1.83 eV*fs, while it becomes almost twice as large for a flattop-like beam resulting in a sinc-like IRF. Therefore, the optimal performance could be achieved for transmission of >4σ of the Gaussian-like beam cross-section. The estimated geometrical transmission of the beamline in the vertical direction, given by the grating open aperture and by the beam divergence, is shown in Fig. 4(a) for the cases of 500 mm long grating and 120 mm long gratings; the corresponding shape of the transmitted cut Gaussian-like beam is shown in inserts at a few points. Operation of the monochromator at constant included angle results unavoidably in a decrease in geometrical transmission at the low photon energy side of operation range; moreover, a substantial increase in divergence towards low photon energies decreases transmission at low photon energies further. One can see that a 500 mm long grating would allow the transmission of >4σ of the Gaussian-like beam cross section for the substantial part of the operation range. In contrast, with an operational 120 mm long grating, the transmission is below 1.7σ over the full operation range. A decreased transmission not only decreases flux on the sample, it results in less optimal time-bandwidth product, as can be seen in Fig. 4(b), where the estimated time-bandwidth product is presented for the cases of ideal optics and design or real optics. For real gratings, the time-bandwidth product would increase compared to that for ideal gratings, since the resolution would degrade due to aberrations while the pulse stretching would stay the same. That is why the ultimate goal would be to keep all kind of aberrations below the IRF of the ideal grating. In the case of installed gratings of the SASE3 monochromator, the time-bandwidth product increases towards higher photon energy ranges of operation of each premirror as a consequence of increasing influence of slope error; the time-bandwidth product is smaller in the case of 150 l/mm grating due to lower slope error and lower VLS miscut. Limiting the grating illumination by closing the aperture before the grating in order to reduce pulse stretching would improve the time-bandwidth product. Finally, to characterize the time-bandwidth product at experiment, the measured resolution should be taken into account instead of calculated one.

**4.2. Photon energy resolution and temporal resolution: performance**

The estimations of resolving power and pulse stretching for the 50 l/mm and 150 l/mm 120 mm long installed gratings in comparison with the 50 l/mm 500 mm long design grating are presented in Fig. 5. In addition to operation in 1$^{st}$ diffraction order, operation in 2$^{nd}$ diffraction order is considered for the case of the 50 l/mm 120 mm long grating because of the regular usage of this mode for experiments demanding strong suppression of higher FEL harmonics. The estimations were done taking into account slope error and VLS miscut affecting the resolution, as described in the previous subsection. The 50 l/mm 500 mm long design grating was meant to achieve about 10000 resolving power. Fig.5 shows that such a resolving power about 10000 should be achievable with the 150 l/mm 120 mm long installed grating, although accompanied with lower transmission compared to the design grating. The 150 l/mm grating is optimized for the experiments demanding high photon energy resolution rather than high temporal resolution. On the other hand, the 50 l/mm 120 mm long grating operational in 1$^{st}$ diffraction order should allow for the moderate resolving power of 2000 - 5000 along with smaller pulse stretching in the range of few to few tens of femtoseconds RMS. This grating is optimized for experiments demanding high temporal resolution. Although, in the low photon energy range, the pulse stretching by this grating is still quite noticeable compared to typical FEL pulse duration. For experiments, demanding shorter pulses, there is a possibility to reduce the pulse stretching by closing the aperture located prior to the grating, as shown in Fig. 3 (b). Finally, operation of the 50 l/mm 120 mm long grating in 2nd diffraction order allows to increase the resolving power compared to operation in 1st diffraction order, although this increase is limited by the VLS miscut.

The photon energy resolution of the two installed gratings was evaluated and optimized using absorption resonances at the Ne K-edge, as described in Section 3. A typical transmission spectrum at Ne K-edge is shown in Fig. 6(a). To optimize the resolution, the longitudinal focusing by the LE premirror has been optimized by changing its angle of incidence. The results of this optimization are shown in Fig. 6 (b), where the open circles represent the resolution, estimated from measurements as described above, versus angle of incidence on the LE premirror. The upper scale shows the longitudinal focus position of the premirror, corresponding to the angle of incidence. The solid circles represent the results of wavefront propagation simulations. In the case of the 50 l/mm grating, the grating was assumed to be plane (the radius of curvature is equal to infinity), and the optimal focusing distance of the premirror corresponds to the focal distance of the monochromator. However, the radius of curvature of -198 km of the 150 l/mm grating (Table 2) shifts the focus downstream, therefore the premirror should be aligned in a way to compensate for this shift. The results of wavefront propagation show, that around 867 eV, the focusing of the LE premirror should be moved upstream by ~10 m, achievable by decreasing the angle of incidence by ~1.4 mrad. Similar results have been obtained analytically. Furthermore, the influence of residual curvature of the grating on the longitudinal focusing is photon energy-dependent; the compensation by the premirror angle becomes more difficult at low photon energies. Thus, at 500 eV one has to move the focusing by the premirror upstream by additional several meters, which becomes difficult to achieve by adjusting the angle of the premirror because of the geometrical constraints of the monochromator. The dependence of measured resolution on the angle of incidence of the LE premirror is very similar to those estimated by wavefront propagation. The measurements in the case of the 150 l/mm grating are limited by the natural bandwidth of Ne lines; the minimal resolution that could be measured with this method was 130 meV, therefore the optimal angle of incidence on the LE premirror was chosen based on the wavefront propagation simulations for this grating.

The optimized resolving power around 867 eV, measured as described above, is ~3300 for the 50 l/mm grating operating in 1st diffraction order and ~4900 for the 50 l/mm grating operating in 2nd diffraction order. For the 150 l/mm grating, the measured resolving power of 6700 is limited by the measurement technique. Yet, the resolving power of the 150 l/mm grating operating in 1st diffraction order has been proven to be > 10000 in the range 500 - 950 eV. It was done by the measurements of combined resolution with hRIXS spectrometer; these measurements showed the resolving power of ~10000 at both O K-edge (~530 eV) and Cu L-edge (~930 eV) [Schlappa *et al.*, 2022]. For both gratings, the measured resolving power is in a good agreement with analytical estimations (presented in Fig. 5a).

### 4.3. Transmission

The estimated and measured transmission of the beamline operating in the monochromatic mode is shown in Fig. 7(a). The transmission is presented through the exit slits of different widths for different gratings; the slit width is chosen based on the IRF presented in Fig. 7(b): 20 μm for the 50 l/mm 500 mm long design grating, 50 μm for the 150 l/mm 120 mm long installed grating and 100 μm for the 50 l/mm 120 mm long installed grating. Note that, because of larger IRF, the wider slit width is possible to use without deterioration of resolution at lower photon energies where the transmission is low. In Fig. 7(b), the IRF corresponds to the estimated resolution, while, based on the resolution measurements presented above, a broader by a factor of ~1.5 IRF is expected in practice.

There is an excellent agreement between the measurements of transmission and the estimations for the case of operation with the HE premirror. The measured transmission in the case of the LE premirror, performed a year later, is lower than the estimated one, which may be an indication of slow contamination of the beamline optics.

### 5. Towards optimal performance

As it can be seen from the discussion above, provided the grating of ultimate quality can be produced, an increase of the grating length could improve not only transmission but also time-bandwidth product. In practice, approaching the design performance by implementing a longer grating is a very challenging task. First of all, one has to overcome limitations in the grating production. The VLS *b2* parameter became critical for the holographic grating production. In the case of mechanical ruling, since the VLS *b2* parameter can be set to 0 within tolerances, the target mean value should be easily achievable; however, mechanical ruling is currently much less developed, compared to holographic ruling, and it suffers from other problems, such as large statistical deviations from the mean values of the ruling parameters over the grating length. In addition, attention has to be paid to the residual curvature of the grating since it shifts the longitudinal focus from the focal plane. Compensation of such a shift by changing the angle of the premirror (discussed in Section 4.2) is possible only in a limited range. An influence of residual curvature on resolution would increase strongly with an increase in the grating length because of a strong decrease of depth of focus for longer grating, allowing for larger acceptance and tighter focusing at the same time. Similarly, all the other factors influencing the longitudinal focusing would become much more pronounced, and, at some point, critical, with increasing grating length. One of such factors is position of the source in the undulator system. If, hypothetically, the shift of focus could be corrected by introducing adaptive properties to the premirror, the prolongation of the source would lead unavoidably to deterioration of resolution when approaching the ultimate performance. The longitudinal properties of the source could become critical even for smaller gratings sizes in certain XFEL operation modes, such as strong quadratic tapering. More limitations could come due to induced thermal deformations (bump) of the long grating surface in the case of using full power of the XFEL beam [La Civita *et al*., 2014].

## 6. Summary

The SASE3 beamline is equipped with a grating monochromator allowing to reduce photon energy bandwidth at the experiments and to improve limited longitudinal coherence of SASE radiation. The monochromator has been designed with the goal to minimize time-bandwidth product and to transport close to transform-limited pulses. Presently, the monochromator is equipped with two gratings: the low-resolution grating, optimized for time-resolved experiments and allowing for moderate resolving power of about 2000 - 5000 along with pulse stretching of few to few tens of femtoseconds RMS, and the high-resolution grating, reaching resolving power of 10000 at a cost of larger pulse stretching.

**Acknowledgements**    The authors acknowledge the accelerator operation team at DESY in Hamburg, Germany, and the facility staff at European XFEL in Schenefeld, Germany, for the provision of stable XFEL beam operation at the SCS instrument and for their assistance.

**References**

Altarelli, M. *et al.* (2006). Technical Design Report, DESY Report 2006-097.

De Fanis, A., Saito, N., Yoshida, H., Senba, Y., Tamenori, Y., Ohashi, H., Tanaka, H. & Ueda, K. (2002). *Phys. Rev. Lett.* **89**, 243001.

Gerasimova, N. (2018). Technical Report, XFEL.EU TR-2018-001.

Koch, A., Risch, J., Freund, W., Maltezopoulos, T., Planas, M. & Grünert J. (2019). *J. Synchrotron Rad.* **26**, 1489–1495

La Civita, D., Gerasimova, N., Sinn, H. & Vannoni, M. (2014). *Proc. SPIE* **9210**, 921002.

Maltezopoulos, T., Dietrich, F., Freund, W., Jastrow, U. F., Koch, A., Laksman, J., Liu, J., Planas, M., Sorokin, A., Tiedtke. K. & Grünert, J. (2019). *J. Synchrotron Rad*. **26**, 1045-1051.

Noda, H., Namioka, T. & Seya, M. (1974). *J. Opt. Soc. Am.* **64**, 1031-1036.

Samoylova, L., Buzmakov, A., Chubar, O. & Sinn, H. (2016). *J. Appl. Cryst.* **49**, 1347-1355.

Schäfers, F. & Krumrey, M. (1996). BESSY Technischer Bericht TB 201, 1

Schäfers, F. (1996). BESSY Technischer Bericht TB 202, 1

Tschentscher, T., Bressler, C., Grünert, J., Madsen, A., Mancuso, A., Meyer, M., Scherz, A., Sinn, H. & Zastrau, U. (2017). *Appl. Sci.* **7**, 592.

Schneidmiller, E. & Yurkov, M. (2011). Preprint, XFEL.EU TR-2011-006.

Sinn, H., Gaudin, M., Samoylova, L., Trapp, A. & Galasso, G. (2011). Conceptual Design Report, XFEL.EU TR-2011-002.

Sinn, H., Dommach, M., Dong, X., La Civita, D., Samoylova, L., Villanueva, R. & Yang, F. (2012). Technical Design Report, XFEL.EU TR-2012-006.

Schlappa, J. *et al.,* (2022). In preparation.

Villanueva, R. *et al.,* (2022). In preparation.

**Table 1** Optical elements of the SASE3 beamline: mirrors

|  | Horizontally deflecting optics | | Vertically deflecting optics | |
| --- | --- | --- | --- | --- |
|  | M1 / M2, offset mirrors | M5 / M6*, distribution mirrors | M3a / M3b, focusing premirrors | M4, deflection mirror |
| Optical surface size | 850 *mm* x 20 *mm* | 850 *mm* x 20 *mm* | 580 *mm* x 25 *mm* | 500 *mm* x 30 *mm* |
| Shape | plane / adaptive | plane | elliptic cylinder | plane |
| Tangential slope error, *RMS* | 56 / 58 *nrad* | 75 / 50 *nrad* | 63 / 37 *nrad* | 140 *nrad* |
| Tangential height error, *PV* | 1.6 / 1.9 *nm* | 1.9 / 2 *nm* | 2.1 / 2.6 *nm* | 7 *nm* |
| Roughness, *RMS* | 0.14 / 0.13 *nm* | 0.2 / 0.2 *nm* | 0.23 / 0.15 *nm* | 0.15 *nm* |
| Grazing angle of incidence | (9 - 20) *mrad* | 9 *mrad* | 9 / 20 *mrad* | 9, 20 *mrad* |

* Specification

**Table 2** Optical elements of the SASE3 beamline: gratings

| Grating | LR Gr, low-resolution design grating* | LR Gr, low-resolution installed grating | HR Gr, high-resolution installed grating |
|---|---|---|---|
| Tangential radius of curvature | | >300 *km* | -198 *km* |
| Optical surface size | 500 *mm* x 25 *mm* | 120 *mm* x 17 *mm* | 120 *mm* x 20 *mm* |
| Slope error, *RMS* | <50 *nrad* | 200 *nrad* | 130 *nrad* |
| Height error, *PV* | <3 *nm* | 6 *nm* | 3.6 *nm* |
| Roughness, *RMS* | <0.2 *nm* | <0.2 *nm* | <0.6 *nm* |
| Included angle | π - 18 *mrad*, π - 40 *mrad* | | |
| VLS law: | $n(w)=b0+b1*w+b2*w^2+..$ | | |
| Central line density $b0$ | 50 *l/mm* | 50.05 *l/mm* | 150 *l/mm* |
| $b1$ | $1.01*10^{-3}$ *l/mm²* | $1.0*10^{-3}$ *l/mm²* | $3.029*10^{-3}$ *l/mm²* |
| $b2$ | 0 *l/mm³* | $-1.2*10^{-6}$ *l/mm³* | $-8.7*10^{-8}$ *l/mm³* |
| Groove profile | Blazed, 0.1° blaze angle | Laminar, 16 *nm* depth | Laminar, 16 *nm* depth |

* Specification

**Figure 1** Layout of the SASE3 beamline, adapted from [Sinn *et al.*, 2012].

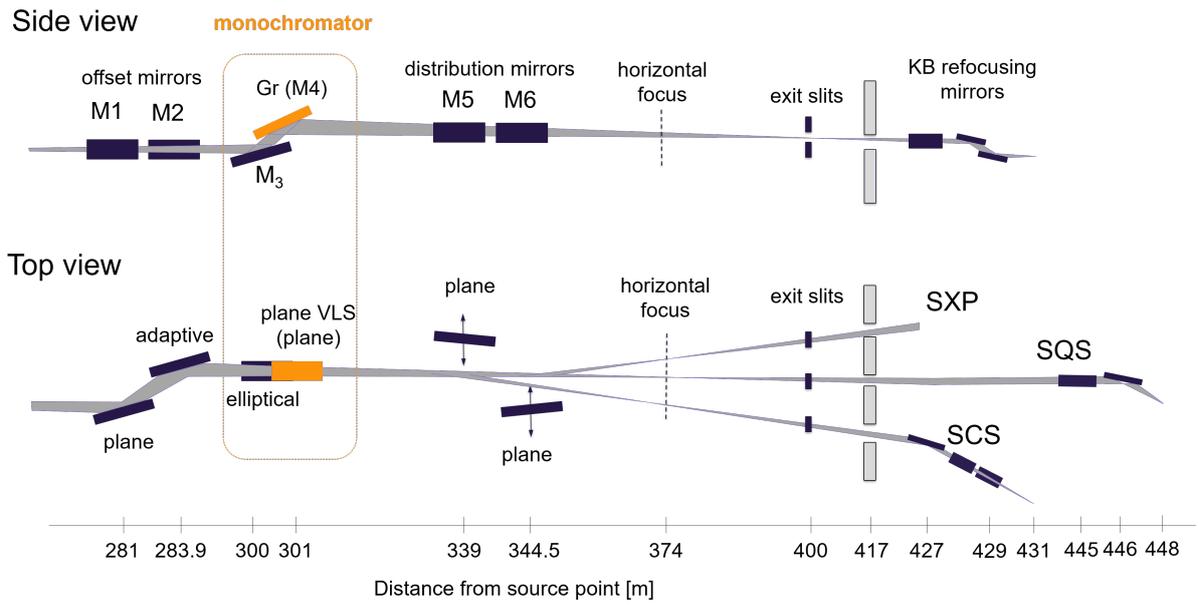

**Figure 2** Influence of figure error and VLS miscut on resolution. (a, c) Analytically estimated resolution of the monochromator operating in 1$^{st}$ diffraction order with the 500 mm long 50 l/mm design grating (slope error 50 nrad RMS, ideal VLS) and with the 120 mm long 50 l/mm grating (slope error 200 nrad RMS, VLS $b2 = -1.2*10^{-6}$ l/mm$^3$; slope error on premirror 50 nrad RMS. (b) Instrument response function of the 500 mm long 50 l/mm design grating operating in 1$^{st}$ diffraction order at 1.5 keV obtained by wavefront propagation simulations for the cases of different magnitudes of slope error on grating; height error / slope error on premirror 3 nm PV / 40 nrad RMS. (d) Analytically estimated tolerances on VLS $b2$ parameter for the 500 mm long 50 l/mm design grating with slope error 50 nrad RMS and for the 120 mm long 50 l/mm grating with slope error 200 nrad RMS. For all panels, solid lines – operation with LE premirror, dashed lines – operation with HE premirror.

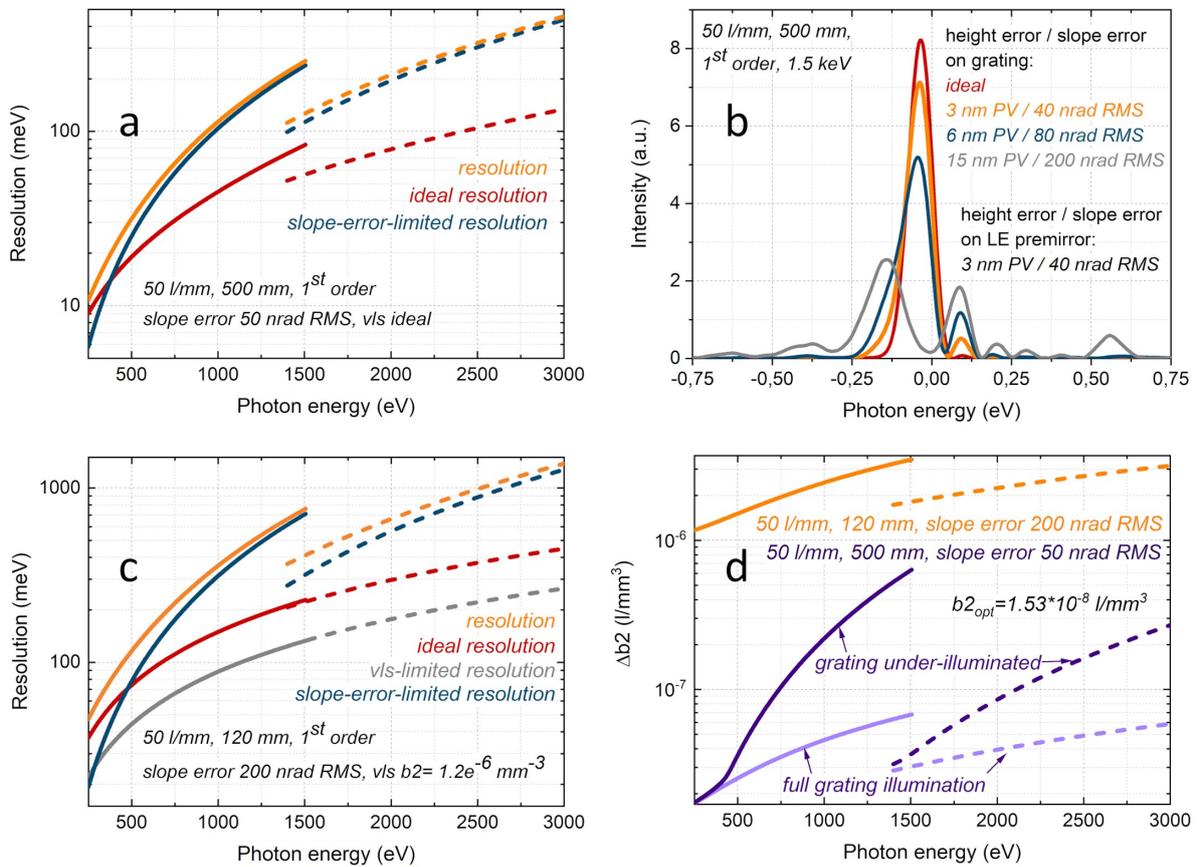

**Figure 3** Influence of grating length on resolution and pulse stretching. (a) Dependence of analytically estimated resolution of the 50 l/mm grating with 50 nrad RMS slope error, operating in 1st diffraction order at 500 eV, on the grating length. Solid circles - ideal VLS, open circles - VLS $b2$ = -1.2*10$^{-6}$ l/mm³. (b) Pulse stretching versus open aperture upstream of the 50 l/mm grating operating in 1st diffraction order: solid line – 290 eV, dashed line – 500 eV, dotted line – 800 eV. The stars correspond to 120 mm length of grating illumination.

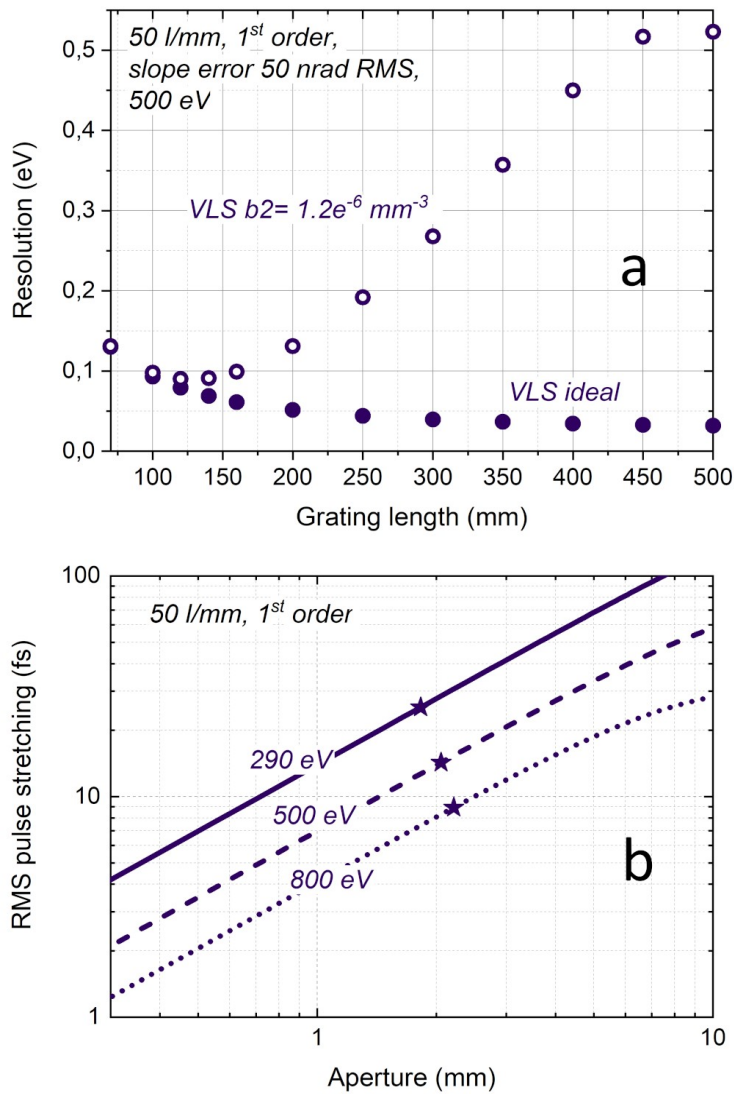

**Figure 4** Time-bandwidth product. (a) Analytically estimated optics transmission in sigma of Gaussian-like beam cross-section; the inserts show the shape of the beam transmitted by the monochromator. (b) Analytically estimated time-bandwidth product in the case of ideal optics with a perfect slope and a perfect VLS law, as well as in the case of a design grating and installed gratings. Operation in 1$^{st}$ diffraction order; solid lines – operation with LE premirror, dashed lines – operation with HE premirror.

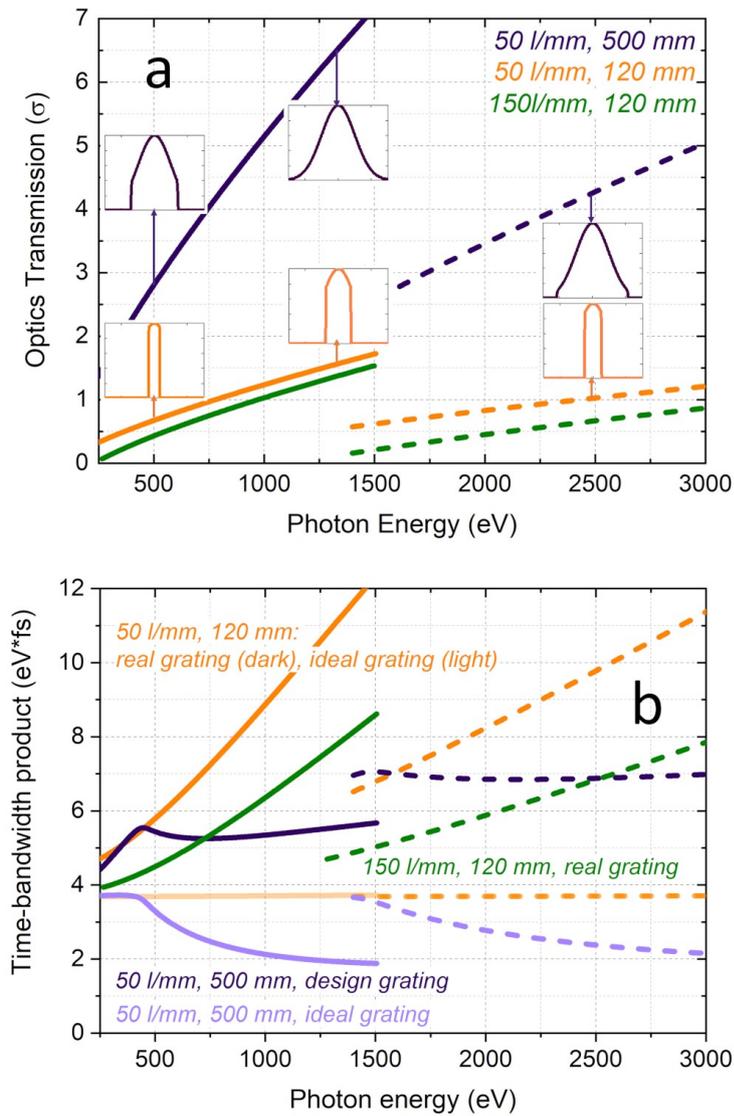

**Figure 5** Monochromator performance. Analytically estimated resolving power (a) and pulse stretching due to the grating (b) for the 50 l/mm 500 mm long design grating operating in 1$^{st}$ diffraction order, for the 50 l/mm 120 mm long installed grating operating in 1$^{st}$ and 2$^{nd}$ diffraction orders, and for the 150 l/mm 120 mm long installed grating operating in 1$^{st}$ diffraction order. Solid lines – operation with LE premirror, dashed lines – operation with HE premirror.

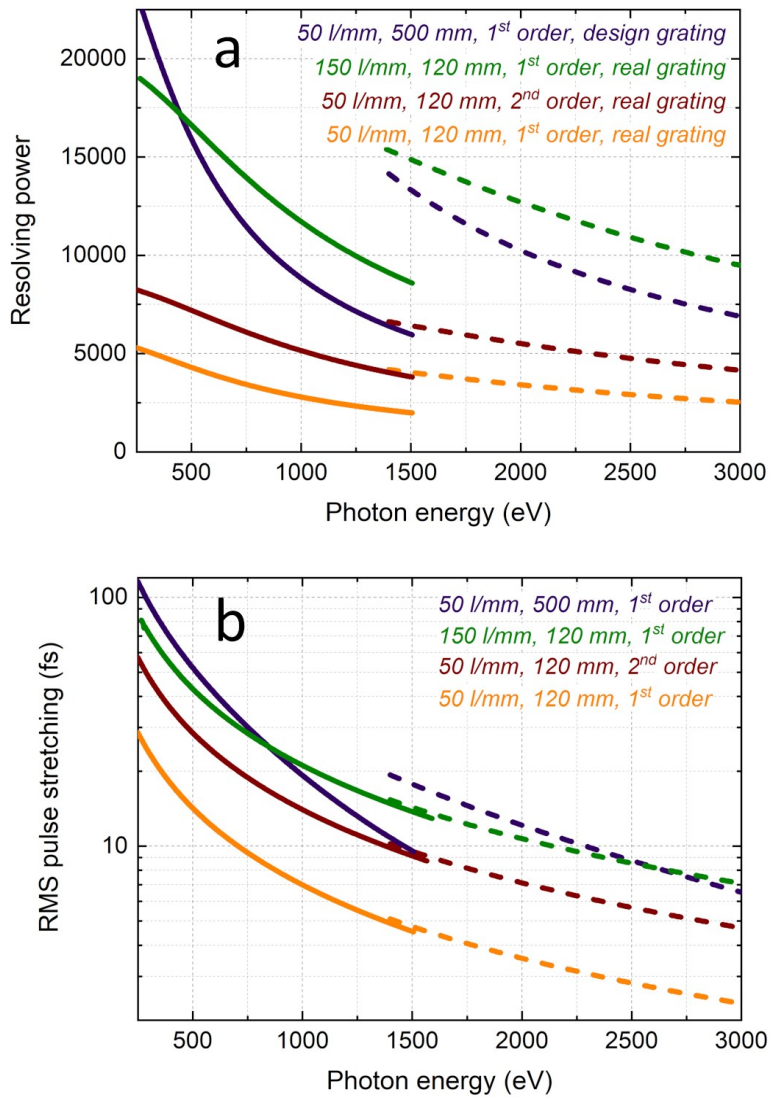

**Figure 6** Resolution measurements and longitudinal focusing. (a) Transmission spectrum of Ne around K-edge; circles represent measured transmission spectra, grey lines – modelled transmission, red lines – resulting convolution of modelled transmission with the monochromator IRF. (b) Resolution optimization at Ne K-edge by aligning the angle of incidence of LE premirror: orange circles – 50 l/mm grating operating in 1st diffraction order, green circles – 150 l/mm grating operating in 1st diffraction order, open circles - measurements, solid circles – wavefront propagation.

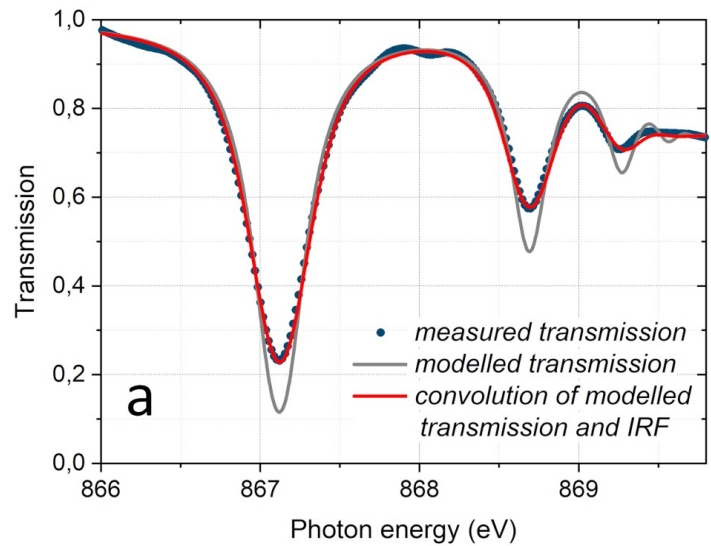

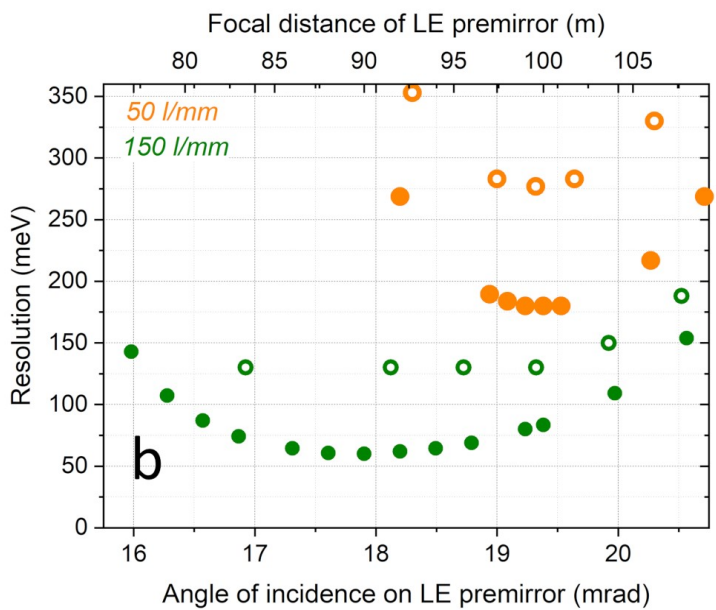

**Figure 7** Transmission. (a) Estimated (lines) and experimentally measured (circles) beamline transmission operating with the 50 l/mm 500 mm long design grating (20 μm exit slit), with the 50 l/mm 120 mm long installed grating (100 μm exit slit), and with the 150 l/mm 120 mm long installed grating (50 μm exit slit). (b) Instrument response function of respective gratings. Solid lines and solid circles– operation with LE premirror, dashed lines and open circles – operation with HE premirror.

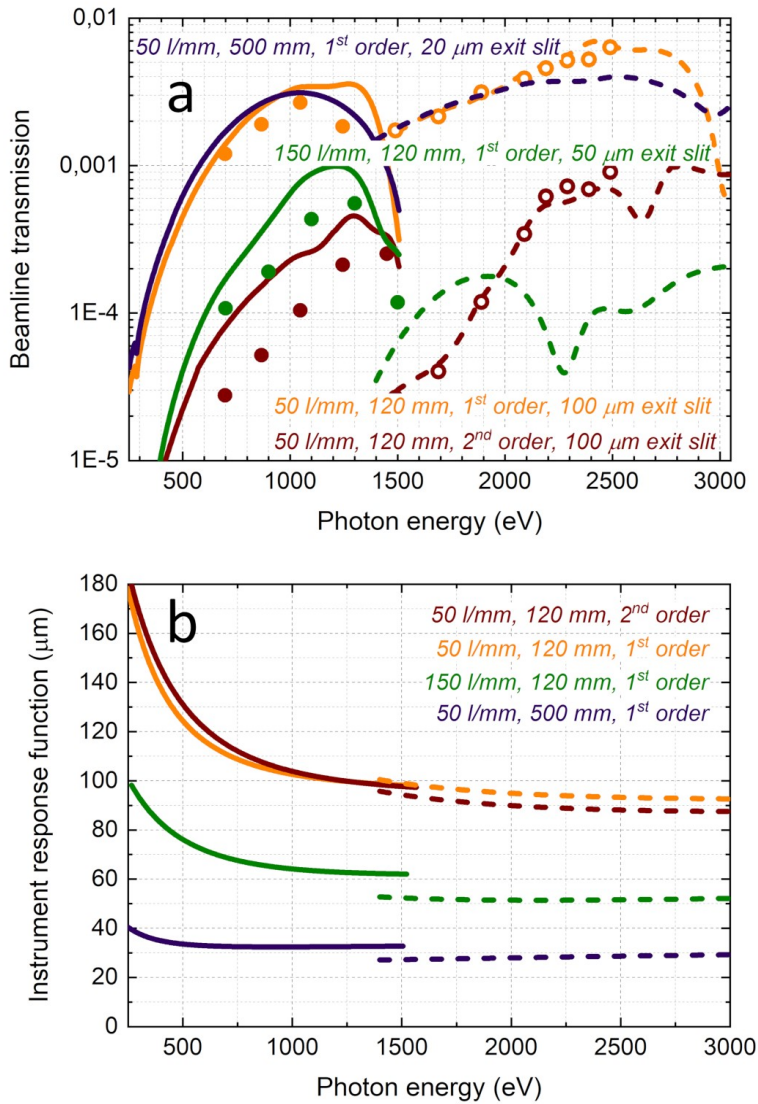